\documentclass[12pt]{amsart}
\usepackage{amssymb}
\newtheorem{claim}{}[section]
\newtheorem{thm}[claim]{Theorem}

\begin{document}
\baselineskip 6.2 truemm
\parindent 1.5 true pc
\newcommand\lan{\langle}
\newcommand\ran{\rangle}
\newcommand\tr{{\text{\rm Tr}}\,}
\newcommand\ot{\otimes}
\newcommand\wt{\widetilde}
\newcommand\join{\vee}
\newcommand\meet{\wedge}
\renewcommand\ker{{\text{\rm Ker}}\,}
\newcommand\im{{\text{\rm Im}}\,}
\newcommand\mc{\mathcal}
\newcommand\transpose{{\text{\rm t}}}
\newcommand\FP{{\mathcal F}({\mathcal P}_n)}
\newcommand\ol{\overline}
\newcommand\JF{{\mathcal J}_{\mathcal F}}
\newcommand\FPtwo{{\mathcal F}({\mathcal P}_2)}
\newcommand\hada{\circledcirc}
\newcommand\id{{\text{\rm id}}}
\newcommand\tp{{\text{\rm tp}}}
\newcommand\pr{\prime}
\newcommand\inte{{\text{\rm int}}\,}
\newcommand\ttt{{\text{\rm t}}}
\newcommand\spa{{\text{\rm span}}\,}
\newcommand\conv{{\text{\rm conv}}\,}
\newcommand\rank{\ {\text{\rm rank of}}\ }
\newcommand\vvv{\mathbb V_{m\meet n}\cap\mathbb V^{m\meet n}}
\newcommand\re{{\text{\rm Re}}\,}
\newcommand\la{\lambda}
\newcommand\msp{\hskip 2pt}
\newcommand\ppt{\mathbb T}

\title{Entangled states with positive partial transposes arising from indecomposable positive linear maps}
\author{Kil-Chan Ha${}^1$, Seung-Hyeok Kye${}^2$ and Young-Sung Park${}^3$}
\thanks{
\indent $ $partially supported by KOSEF\\
\indent ${}^1$Applied Mathematics, Sejong University, Seoul 143-747, KOREA\\
\indent ${}^2$Department of Mathematics, Seoul National
University, Seoul 151-742, KOREA \\
\indent ${}^3$Department of Mathematics, Seoul National
University, Seoul 151-742, KOREA} 
\maketitle

\begin{abstract}
We construct entangled states with positive partial transposes
using indecomposable positive linear maps between matrix algebras.
We also exhibit concrete examples of entangled states with
positive partial transposes arising in this way, and show that
they generate extreme rays in the cone of all positive
semi-definite matrices with positive partial transposes. They also
have Schmidt numbers two.
\end{abstract}
\section{Introduction}

The notion of entanglement in quantum physics has been playing a
key r\^ ole in the quantum information theory and quantum
communication theory during the past decade (see \cite{horo} for a
survey). On the other hand, properties of positive linear maps
between matrix algebras have been studied by operator algebraists
from the sixties \cite{stormer}. It is now turned out that these
two theories are dual each other. For example, the second author
\cite{eom-kye} used various cones of block matrices adapting the
idea of Woronowicz \cite{woronowicz} to characterize the boundary
structures of various kinds of positive linear maps between matrix
algebras. One of these notions for block matrices is turned out to
be nothing but entangled states. On the other hand, people in the
quantum information theory used the notion of positive linear maps
to characterize entangled states. See \cite{horo-1}, \cite{horo-2}
and \cite{terhal_horo} for examples. This duality was used
\cite{terhal} to construct indecomposable positive linear maps
arising from entangled states.

In this paper, we use the duality to construct entangled states
with positive partial transposes from decomposable positive linear
maps which lie on the boundary of the cone of all decomposable
positive maps but belong to the interior of the cone of all
positive linear maps. Every indecomposable positive linear map
gives us such an example of decomposable positive map.

This construction will be applied to a family of indecomposable
positive linear maps in \cite{cho-kye-lee} which are variants of
those maps constructed by Choi in the seventies \cite{choi75},
\cite{choi-lam}. This gives us a family of $3\otimes 3$ entangled
states which generate extreme rays in the cone of all $9\times 9$
positive semi-definite matrices with positive partial transposes.
Finally, we show that these entangled states have Schmidt numbers
two by explicit constructions.

After we explain the duality theory between linear maps and block
matrices in the second section, we will exhibit in the third
section the method to construct entangled states with positive
partial transposes. In the final section, we give concrete
examples of $3\otimes 3$ entangled states obtained in this way,
and examine their properties mentioned above.

Throughout this paper, we will not use bra-ket notation. Every vector will be considered as a column vector. If
$x\in\mathbb C^m$ and $y\in \mathbb C^n$ then $x$ will be considered as an $m\times 1$ matrix, and $y^*$ will be
considered as a $1\times n$ matrix, and so $xy^*$ is an $m\times n$ rank one matrix whose range is generated by $x$
and whose kernel is orthogonal to $y$. For a vector $x$ and a matrix $A$, the notations $\ol x$ and $\ol A$ will
be used for the vector
and matrix, respectively, whose entries are conjugate of the corresponding entries. The notation $\lan\cdot,\cdot \ran$ will
be used for bi-linear pairing. On the other hand, $(\cdot |\cdot )$ will be used for the inner product, which is
sesqui-linear, that is, linear in the first variable and conjugate-linear in the second variable.

The second author is grateful to Professor D.-P. Chi for bringing
his attention to the notion of entanglement.

\section{Duality}

A linear map between $C^*$-algebras is said to be {\sl positive} if it send every positive element
to a positive element. The whole structures of the cone of all positive linear maps
is extremely complicated even in low dimensional matrix algebras and far from being understood completely.
A linear map $\phi:A\to B$ is said to be
{\sl $s$-positive} if the map
$$
\phi_s: M_s(A)\to M_s(B): [a_{ij}]\mapsto [\phi(a_{ij})]
$$
is positive, where $M_s(A)$ is the $C^*$-algebra of all $s\times s$ matrices over $A$. We say that
$\phi$ is {\sl completely positive} if $\phi$ is $s$-positive for every $s=1,2,\dots$.
The transpose map
$$
X\mapsto X^{\ttt},\qquad X\in M_n
$$
is a typical example of a positive linear map which is not completely positive, where $M_n$ denotes the
$C^*$-algebra of all $n\times n$ matrices over the complex field. We say that a linear map
$\phi: M_m\to M_n$ is said to be {\sl $s$-copositive} if $X\mapsto \phi(X^{\ttt})$ is $s$-positive, and
{\sl completely copositive} if it is $s$-copositive for every $s=1,2,\dots$. The cone of all $s$-positive
(respectively $t$-copositive) linear maps
from $A$ into $B$ will be denoted by $\mathbb P_s[A,B]$ (respectively $\mathbb P^t[A,B]$), and just by
$\mathbb P_s$ (respectively $\mathbb P^t$) whenever the domain and range are clear.

From now on, we identify an $m\times n$ matrix $z\in M_{m\times n}$  and a vector
$\wt z\in \mathbb C^n\otimes\mathbb C^m$ as follows: For $z=[z_{ik}]\in
M_{m\times n}$, we define
$$
\begin{aligned}
z_i=&\sum_{k=1}^n z_{ik} e_k\in \mathbb C^n,\qquad i=1,2,\dots, m,\\
\wt z=&\sum_{i=1}^m z_i\otimes e_i\in \mathbb C^n\otimes\mathbb C^m.
\end{aligned}
$$
Then $z\mapsto \wt z$ defines an inner product isomorphism
from $M_{m\times n}$ onto $\mathbb C^n\otimes \mathbb C^m$.
For a given $m\times n$ matrix $z\in M_{m\times n}$, we note that ${\wt z}\, {\wt z}^*$ belongs to
$M_n\otimes M_m$, which is identified with the space $M_m(M_n)$ of all $m\times m$ matrices
whose entries are $n\times n$ matrices. For $A\in M_n\otimes M_m$, we denote by $A^\tau$
the {\sl block transpose} or {\sl partial transpose} of $A$, that is,
$$
\left(\sum_{i,j=1}^m a_{ij}\otimes
e_{ij}\right)^\tau=\sum_{i,j=1}^m a_{ji}\otimes e_{ij}.
$$
Now, we define
$$
\begin{aligned}
\mathbb V_s=&\conv \{\wt z\,{\wt z}^*\in M_n\otimes M_m: \rank z\le s\},\\
\mathbb V^s=&\conv \{(\wt z\,{\wt z}^*)^\tau\in M_n\otimes M_m: \rank
z\le s\},
\end{aligned}
$$
for $s=1,2,\dots,m\meet n$, where $\conv X$ means the convex
set generated by $X$, and $m\meet n$ denotes the minimum of $m$ and $n$. It is clear that $\mathbb V_{m\meet n}$
coincides with the cone of all positive semi-definite $mn\times mn$ matrices. It is easily seen that
$\mathbb V_1=\mathbb V^1$. We also have the following chains
$$
\mathbb V_1\subset\mathbb V_2\subset\cdots\subset\mathbb V_{m\meet n},\qquad
\mathbb V^1\subset\mathbb V^2\subset\cdots\subset\mathbb V^{m\meet n}
$$
of inclusions.
We note that a block matrix $A\in M_n\otimes M_m$ with $\tr A=1$ represents a {\sl separable} state if and only if
$A\in\mathbb V_1$. Therefore, we see that a block matrix $A\in M_n\otimes M_m$ with $\tr A=1$
represents an {\sl entangled} state if and only if $A\notin \mathbb V_1$.
The minimum number $s$ with $A\in\mathbb V_s$ is the {\sl Schmidt number} of $A\in M_n\otimes M_m$, in the language of
quantum information theory.

Motivated by the work of Woronowicz \cite{woronowicz} (see also \cite{itoh}, \cite{stormer82}),
we have considered in \cite{eom-kye} the bi-linear pairing between
$M_n\otimes M_m$ and the space ${\mathcal L}(M_m, M_n)$ of all linear
maps from $M_m$ into $M_n$, given by
\begin{equation}\label{definition}
\lan A,\phi\ran =\tr \left[ \left(\sum_{i,j=1}^m
\phi(e_{ij}) \otimes e_{ij}\right) A^\ttt \right]
=\sum_{i,j=1}^m\lan\phi(e_{ij}),a_{ij}\ran,
\end{equation}
for $A=\sum_{i,j=1}^m a_{ij}\ot e_{ij}\in M_n\ot M_m$ and
$\phi\in{\mathcal L}(M_m, M_n)$, where the
bi-linear form in the right-side is given by $\lan a,b\ran=\tr (ba^\ttt )$
for $a,b\in M_n$.
This is equivalent to define
$$
\lan y\otimes x,\phi\ran
=\tr (\phi(x)y^\ttt),\qquad x\in M_m,\ y\in M_n.
$$
In this duality, the pairs
$$
(\mathbb V_s,\ \mathbb P_s),\qquad
(\mathbb V^t,\ \mathbb P^t),\qquad
(\mathbb V_s\cap\mathbb V^t,\ \mathbb P_s+\mathbb P^t)
$$
are dual each other, in the sense that
$$
\begin{aligned}
A\in \mathbb V_s\ &\Longleftrightarrow \
  \lan A,\phi\ran \ge 0\ {\text{\rm for each}}\ \phi\in\mathbb
P_s,\\ \phi\in \mathbb P_s\ &\Longleftrightarrow \
  \lan A,\phi\ran \ge 0\ {\text{\rm for each}}\ A\in\mathbb V_s,
\end{aligned}
$$
and similarly for others. These dualities are turned out to be the
basic key idea in the various characterizations of entangled
states studied in quantum information theory.

Now, we restrict ourselves to the duality between the cone $\mathbb V_{m\meet n}\cap\mathbb V^{m\meet n}$
and the cone $\mathbb P_{m\meet n}+\mathbb P^{m\meet n}$. A linear maps in
the cone
$$
\mathbb D:=\mathbb P_{m\meet n}+\mathbb P^{m\meet n}
$$
is said to be {\sl decomposable}, that is, a linear map is said to be decomposable if it is the sum of
a completely positive linear map and a completely copositive linear map. Every decomposable
map is positive, but the converse is not true. There are many examples of indecomposable positive linear maps
in the literature \cite{cho-kye-lee}, \cite{choi75}, \cite{ha}, \cite{ha-1}, \cite{kim-kye}, \cite{kye-22},
\cite{osaka}, \cite{robertson}, \cite{stormer82}, \cite{tomiyama}, \cite{tang}, \cite{terhal}. Since the cone
$\mathbb V_{m\meet n}$ consists of all positive semi-definite matrices, the cone
$$
\ppt:=\mathbb V_{m\meet n}\cap\mathbb V^{m\meet n}
$$
consists of all positive semi-definite matrices whose block transposes are also positive semi-definite, or
positive semi-definite matrices with {\sl positive partial transposes} in the language of
quantum information theory. The duality between two cones $\mathbb D$
and $\ppt$ is summarized by
$$
\begin{aligned}
A\in \ppt\ &\Longleftrightarrow \ \lan A,\phi\ran \ge 0\ {\text{\rm for each}}\ \phi\in\mathbb D,\\
\phi\in \mathbb D\ &\Longleftrightarrow \  \lan A,\phi\ran \ge 0\ {\text{\rm for each}}\ A\in\ppt,
\end{aligned}
$$
for $A\in M_n\otimes M_m$ and $\phi\in{\mathcal L}(M_m,M_n)$.

Now, we proceed to characterize the boundary structures of the cones $\mathbb D$ and $\ppt$.
For a subset $X$ of $\mathbb D$, we define
$$
X^\prime =\{A\in\ppt: \lan A,\phi\ran=0,\ {\text{\rm for every}}\ \phi\in X\}.
$$
Similarly, we also define $Y^\prime$ for a subset $Y$ of $\ppt$ by
$$
Y^\prime =\{\phi\in\mathbb D: \lan A,\phi\ran=0,\ {\text{\rm for every}}\ A\in Y\}.
$$
It is easy to see that $X^\prime$ is a face of $\ppt$, and every exposed face of $\ppt$ arises in this way.
We also note that if $\phi\in\mathbb D$ is an interior point of a face $F$ of $\mathbb D$ then $F^\prime=\{\phi\}^\prime$.
The set $\{\phi\}^\prime$ will be written by $\phi^\prime$.
Dual results also hold, of course. This kind of duality holds for much more general situations, and
was used to characterize maximal faces of the cones $\mathbb P_s[M_m,M_n]$ for
$s=1,2,\dots,m\meet n$ \cite{kye-canad}, \cite{kye-korean}.

Next, we review the intrinsic characterization of faces of the cone $\mathbb D$ as
was in \cite{kye-decom}, which was motivated
to find out all faces of the cone $\mathbb P_1[M_2,M_2]$ of all positive linear maps
between $2\times 2$ matrices \cite{byeon-kye}, \cite{kye-2by2_II}.
For a finite subset ${\mathcal V}=\{V_1, V_2,\dots, V_\nu\}$ of
$M_{m\times n}$, we define the linear maps $\phi_\mathcal V$ and
$\phi^\mathcal V$ from $M_m$ into $M_n$ by
\begin{equation}\label{kraus-rep}
\begin{aligned}
&\phi_\mathcal V: X\mapsto \sum_{i=1}^\nu V_i^*XV_i,\qquad X\in
M_m,\\
&\phi^\mathcal V: X\mapsto \sum_{i=1}^\nu V_i^*X^{\ttt}V_i,\qquad
X\in M_m.
\end{aligned}
\end{equation}
We also denote by $\phi_V=\phi_{\{V\}}$ and $\phi^V=\phi^{\{V\}}$.
It is well-known \cite{choi75-10}, \cite{kraus} that every completely positive
(respectively completely copositive) linear map from $M_m$ into $M_n$ is of the form
$\phi_\mathcal V$ (respectively $\phi^\mathcal V$). Every linear map $\phi:M_m\to M_n$ corresponds to
a block matrix in $M_n\otimes M_m=M_m(M_n)$ by
\begin{equation}\label{choi-rep}
\phi\mapsto [\phi(e_{ij})]_{i,j=1,2,\dots,m}.
\end{equation}
It is also well known \cite{choi75-10} that a linear map $\phi:M_m\to M_n$ is completely positive if and only if
it is $(m\meet n)$-positive if and only if the corresponding matrix in (\ref{choi-rep}) is
positive semi-definite.
For a subspace $E$ of $M_{m\times n}$, we define
\begin{equation}\label{phi_E}
\begin{aligned}
\Phi_E=&\{\phi_\mathcal V\in\mathbb P_{m\meet n}[M_m,M_n]: \spa{\mathcal
V}\subset E\}\\
\Phi^E=&\{\phi_\mathcal V\in\mathbb P^{m\meet n}[M_m,M_n]: \spa{\mathcal
V}\subset E\},
\end{aligned}
\end{equation}
where $\spa{\mathcal V}$ denotes the span of the set $\mathcal V$.
We have shown in \cite{kye-cambridge} that the correspondences
$$
\begin{aligned}
E\mapsto \Phi_E &: {\mathcal
E}(M_{m\times n})\longleftrightarrow{\mathcal F}(\mathbb P_{m\meet n}[M_m,M_n])
\\
E\mapsto \Phi^E &: {\mathcal
E}(M_{m\times n})\longleftrightarrow{\mathcal F}(\mathbb P^{m\meet n}[M_m,M_n])
\end{aligned}
$$
give rise to lattice isomorphisms from the lattice ${\mathcal E}(M_{m\times n})$ of all
subspaces of the vector space $M_{m\times n}$ onto the lattice ${\mathcal F}(\mathbb P_{m\meet n}[M_m,M_n])$
(respectively ${\mathcal F}(\mathbb P^{m\meet n}[M_m,M_n])$)
of all faces of the convex cone $\mathbb P_{m\meet n}[M_m,M_n]$ (respectively $\mathbb P^{m\meet n}[M_m,M_n]$).

Let $C$ be the convex hull of the cones
$C_1$ and $C_2$. If $F$ is a face of $C$ then it is easy to see that $F\cap C_1$ and $F\cap C_2$ are faces of $C_1$ and
$C_2$, respectively, and $F$ is the convex hull of $F\cap C_1$ and $F\cap C_2$. This is immediately applied to
characterize faces of the cone $\mathbb D$ which is the convex hull of the cones $\mathbb P_{m\meet n}$
and $\mathbb P^{m\meet n}$.

For a given face $F$ of $\mathbb D$, we see that $F\cap \mathbb P_{m\meet
n}$ is a face of $\mathbb P_{m\meet n}$, and so it is of the form $\Phi_D$ for a
subspace $D\in {\mathcal E}(M_{m\times n})$. Similarly,
$F\cap \mathbb P^{m\meet n}= \Phi^E$ for a subspace $E\in {\mathcal
E}(M_{m\times n})$. Therefore, we see that every face of
$\mathbb D$ is of the form
\begin{equation}\label{face-form}
\sigma(D,E):=\conv \{\Phi_D,\Phi^E\},
\end{equation}
for $D,E\in {\mathcal E}(M_{m\times n})$. If we assume the following condition
$$
\sigma(D,E)\cap \mathbb P_{m\meet n}= \Phi_D,\qquad
\sigma(D,E)\cap \mathbb P^{m\meet n}= \Phi^E
$$
then it is clear that every face of $\mathbb D$ is uniquely expressed as in
(\ref{face-form}). It seems to be very difficult in general to determine all the pairs of subspaces
which give rise to faces of $\mathbb D$. See \cite{byeon-kye} for the simplest case $m=n=2$.

\section{Construction}

Let $C$ be a convex set. Recall that a point $x\in C$ is said to be an {\sl interior} point of $C$ if
for every $y\in C$ there exists $t>1$ such that $(1-t)y+tx\in C$. A point $x\in C$ is said to be a {\sl boundary}
point if it is not an interior point. Every nonempty convex set has an interior point. Let $y_0\in C$ be a fixed
interior point. Then it is easy to see that a point $x\in C$ is an interior point if and only if
there is $t>1$ such that $(1-t)y_0+tx\in C$.
The set of all interior points (respectively boundary points) of $C$ will be denoted by $\inte C$
(respectively $\partial C$). It is known that the interior of the convex set $\Phi_E$ in (\ref{phi_E}) is given by
$$
\inte \Phi_E=
\{\phi_\mathcal V\in\mathbb P_{m\meet n}[M_m,M_n]: \spa{\mathcal
V}= E\},
$$
for example, and similarly for the convex set $\Phi^E$.
The trace map $A\mapsto (\tr A)I_n$ from $M_m$ into $M_n$, where $I_n$ is the $n\times n$ identity matrix, is a typical
example of an interior point of the cones $\mathbb P_s[M_m,M_n]$ and $\mathbb P^t[M_m,M_n]$ for $s,t=1,2,\dots,
m\meet n$. By an abuse of notations, we will denote this map by ${\text{\rm Tr}}$, again.

Suppose that we are given an example of an indecomposable positive linear map $\phi:M_m\to M_n$. If we define
$$
\alpha=\sup\{t\in\mathbb R: \phi_t:=(1-t)\tr +t\phi\in\mathbb D\},
$$
then, by the above discussion, we see that
$$
\phi_\alpha:=(1-\alpha)\tr +\alpha\phi
$$
is a boundary point of the cone $\mathbb D$, but
is an interior point of the cone $\mathbb P_1$ of all positive linear maps. We recall that a convex set
is partitioned into the family of interiors of the faces. Therefore, the boundary point $\phi_\alpha$ determines
a proper face $\sigma(D,E)$ of $\mathbb D$ whose interior contains $\phi_\alpha$.
Since $\sigma (D,E)$ is a convex subset of $\mathbb P_1$, we have the two cases:
$$
\inte\sigma(D,E)\subset \inte\mathbb P_1
\qquad {\text{\rm or}}\qquad
\sigma(D,E)\subset\partial \mathbb P_1.
$$
The above construction gives us a face $\sigma(D,E)$ of $\mathbb D$ whose
interior is contained in the interior of $\mathbb P_1$.

\begin{thm}\label{main-theorem}
If $(D,E)$ is a pair of spaces of $m\times n$ matrices which gives rise to a proper face $\sigma(D,E)$ of $\mathbb D$
with $\inte\sigma(D,E)\subset \inte\mathbb P_1$ then every nonzero
element $A$ of the face $\sigma(D,E)^\prime$ of $\ppt$
belongs to $\ppt\setminus \mathbb V_1$.
\end{thm}

\begin{proof}
Take an interior point $\phi$ of $\sigma(D,E)$. Then it is also an interior point of the cone $\mathbb P_1$.
Therefore, there is $t>1$ such that
$$
\psi:=(1-t)\tr +t\phi\in\mathbb P_1.
$$
Since $\tr$ is an interior point of the cone $\mathbb D$ and $A\neq 0$, we have $\lan A,\tr\ran >0$.
Furthermore, we have $\lan A,\phi\ran=0$, since $A\in\sigma(D,E)^\prime$.
Therefore, we have
$$
\lan A,\psi\ran=(1-t)\lan A,\tr\ran+t\lan A,\phi\ran=(1-t)\lan A,\tr\ran <0.
$$
This shows that $A\notin \mathbb V_1$ from the duality between $\mathbb P_1$ and $\mathbb V_1$.
\end{proof}

With a single example $\phi_\alpha\in\partial \mathbb D\cap\inte\mathbb P_1$, it is not so easy to
determine the subspaces $D$ and $E$ of
$M_{m\times n}$ so that $\phi_\alpha\in\inte\sigma(D,E)$.
But, it is easy to determine the cone $\sigma(D,E)^\prime$ which coincides with $(\phi_\alpha)^\prime$,
since $\phi_\alpha$ is an interior point of $\sigma(D,E)$.
To do this, we write $\phi_\alpha$ in the form (\ref{kraus-rep}).
We first consider the matrix representation (\ref{choi-rep}) to get the block matrix
$B=[\phi_\alpha(e_{ij})]\in M_n\otimes M_m$. Since $\phi_\alpha$ is decomposable, it is possible to write as $B=P+Q$, where both $P$
and $Q^\tau$ are positive semi-definite. We note that the map $\phi_V$ corresponds to
the block matrix ${\wt{\ol V}}\,{\wt{\ol V}}^*$ by a direct calculation. Therefore, if we write
$$
P=\sum_i {\wt a_i}\,{\wt a_i}^*,\qquad Q^\tau=\sum_j {\wt b_j}\,{\wt b_j}^*,
$$
using the spectral decomposition of $P$ and $Q^\tau$, where $a_i,b_j\in M_{m\times n}$, then it is
possible to write as
$$
\phi_\alpha=\phi_{\mathcal V}+\phi^{\mathcal W}
$$
where $\mathcal V$ and $\mathcal W$ are sets of $m\times n$ matrices.
Note that every
$A\in \sigma(D,E)^\prime$ belongs to $\ppt\subset\mathbb V_{m\meet n}$, which is nothing but the cone of all
positive semi-definite matrices. Therefore, we can write $A=\sum_i {\wt x_i}\,{\wt x_i}^*$ with $x_i\in M_{m\times n}$.
If $V\in D$ then we have
$$
\lan {\wt x_i}\,{\wt x_i}^*, \phi_V\ran
=\lan {\wt x_i}\,{\wt x_i}^*,\wt{\ol V}\, \wt{\ol V}^*\ran
=|(x_i|V)|^2
$$
as was seen in \cite{kye-decom}. Therefore, we see that
$$
0=\lan A,\phi_V\ran =\sum_i |(x_i|V)|^2,
$$
and so $x_i\perp V$. This is true for every $V\in D$, and we see that $x_i\in D^\perp$ for each $i$. Therefore, we have
$$
A=\sum_i \wt x_i\wt x_i^*\in (\phi_{\mathcal V})^\prime\ \Longleftrightarrow\ \spa\{x_i\}\perp\spa{\mathcal V}
$$
and similarly for $(\phi^{\mathcal W})^\prime$. Finally, we note that the relation
$$
(\phi_{\mathcal V})^\prime\cap(\phi^{\mathcal W})^\prime
\subset (\phi_{\mathcal V}+\phi^{\mathcal W})^\prime
=\sigma(D,E)^\prime
=(\Phi_D)^\prime\cap (\Phi^E)^\prime
\subset (\phi_{\mathcal V})^\prime\cap(\phi^{\mathcal W})^\prime
$$
shows that the above sets coincide,
where the first inclusion comes out from the identity $\lan A,\phi_{\mathcal V}+\phi^{\mathcal W}\ran
=\lan A,\phi_{\mathcal V}\ran+\lan A,\phi^{\mathcal W}\ran$, and the last inclusion follows from
$\phi_{\mathcal V}\in\Phi_D$ and $\phi_{\mathcal W}\in\Phi^E$.

\section{Examples}

We begin with the map $\Phi[a,b,c]:M_3\to M_3$ defined by
$$
\Phi[a,b,c]:
x \mapsto
\left(\begin{matrix} ax_{11}+bx_{22}+cx_{33}  &0  &0\\
0 &ax_{22}+bx_{33}+cx_{11} &0\\
0&0&ax_{33}+bx_{11}+cx_{22}\end{matrix}\right)
- x
$$
for $x=(x_{ij})\in M_3$, as was studied in \cite{cho-kye-lee}. Recall that $\Phi[2,0,1]$ is
an example \cite{choi-lam} of an extremal positive linear map which is not decomposable.
It was shown that $\Phi[a,b,c]$ is positive if and only if
$$
a\ge 1,\qquad a+b+c\ge 3,\qquad 1\le a\le 2\ \Longrightarrow bc\ge(2-a)^2,
$$
and decomposable if and only if
$$
a\ge 1,\qquad 1\le a\le 3\ \Longrightarrow bc\ge\left(\dfrac{3-a}2\right)^2.
$$
Therefore, every $\Phi[a,b,c]$ with the condition
$$
1<a<3,\qquad 4bc=(3-a)^2
$$
gives rise to an element of $\partial\mathbb D\cap\inte\mathbb P_1$, whenever
$b\neq c$.
Furthermore, we have a decomposition
$$
\Phi[a,b,c]=\dfrac{a-1}2\Phi[3,0,0]+\dfrac{3-a}2\Phi\left[1,\sqrt{\dfrac bc},\sqrt{\dfrac cb}\right]
$$
into the sum of a completely positive map and a completely copositive map. If we fix $b$ and $c$, then
we see that the family $\{\Phi[a,b,c]: 1 \le a \le3\}$ is a line segment, and so it suffices to consider
the map $\Phi[2,b,c]$. We  also see that
$$
\Phi[3,0,0]=\phi_{V_1}+\phi_{V_2}+\phi_{V_3}
$$
with
$$
V_1=\left(\begin{matrix}1&\cdot&\cdot\\ \cdot&-1&\cdot\\ \cdot&\cdot&\cdot\end{matrix}\right),\quad
V_2=\left(\begin{matrix}\cdot&\cdot&\cdot\\ \cdot&1&\cdot\\ \cdot&\cdot&-1\end{matrix}\right),\quad
V_3=\left(\begin{matrix}-1&\cdot&\cdot\\ \cdot&\cdot&\cdot\\ \cdot&\cdot&1\end{matrix}\right),
$$
where $\cdot$ denotes $0$, and
$$
\Phi\left[1,\sqrt{\dfrac bc},\sqrt{\dfrac cb}\right]
=\phi^{W_1}+\phi^{W_2}+\phi^{W_3}
$$
with
$$
W_1=\left(\begin{matrix} \cdot&\mu&\cdot\\ -\la&\cdot&\cdot\\ \cdot&\cdot&\cdot\end{matrix}\right),\qquad
W_2=\left(\begin{matrix} \cdot&\cdot&\cdot\\ \cdot&\cdot&\mu\\ \cdot&-\la&\cdot\end{matrix}\right),\qquad
W_3=\left(\begin{matrix} \cdot&\cdot&-\la\\ \cdot&\cdot&\cdot\\ \mu&\cdot&\cdot\end{matrix}\right),
$$
where $\lambda=\left (\dfrac bc \right )^{1/4}$ and
$\mu=\left (\dfrac cb \right )^{1/4}$, and so $\la\mu=1$ and $\la\neq 1$.

To find an element of $\Phi[2,b,c]^\prime$, we have to consider the orthogonal complement of the space of $3\times 3$
matrices spanned by $\{V_i,W_i:i=1,2,3\}$. To do this,
we write
$$
x=\displaystyle\left(\begin{matrix} 1&\cdot&\cdot\\ \cdot&1&\cdot\\ \cdot&\cdot&1\end{matrix}\right),\quad
y_1=\left(\begin{matrix} \cdot&\la&\cdot\\ \mu&\cdot&\cdot\\ \cdot&\cdot&\cdot\end{matrix}\right),\quad
y_2=\left(\begin{matrix} \cdot&\cdot&\cdot\\ \cdot&\cdot&\la\\ \cdot&\mu&\cdot\end{matrix}\right),\quad
y_3=\left(\begin{matrix} \cdot&\cdot&\mu\\ \cdot&\cdot&\cdot\\ \la&\cdot&\cdot\end{matrix}\right),
$$
where $\la\mu=1$, $\lambda\neq 1$. Then, it is immediate that
$$
\begin{aligned}
A&=\wt x\wt x^*+\sum_{i=1}^3\wt y_i\wt y_i^*\\
&=
\left(
\begin{array}{ccccccccccc}
1     &\cdot   &\cdot  &&\cdot  &1     &\cdot   &&\cdot   &\cdot  &1     \\
\cdot   &\la^2 &\cdot    &&1    &\cdot   &\cdot &&\cdot &\cdot     &\cdot   \\
\cdot  &\cdot    &\mu^2 &&\cdot &\cdot  &\cdot    &&1    &\cdot &\cdot  \\
\\
\cdot  &1    &\cdot &&\mu^2 &\cdot  &\cdot    &&\cdot    &\cdot &\cdot  \\
1     &\cdot   &\cdot  &&\cdot  &1     &\cdot   &&\cdot   &\cdot  &1     \\
\cdot   &\cdot &\cdot    &&\cdot    &\cdot   &\la^2 &&\cdot &1    &\cdot   \\
\\
\cdot   &\cdot &1    &&\cdot    &\cdot   &\cdot &&\la^2 &\cdot    &\cdot   \\
\cdot  &\cdot    &\cdot &&\cdot &\cdot  &1    &&\cdot    &\mu^2 &\cdot  \\
1     &\cdot   &\cdot  &&\cdot  &1     &\cdot   &&\cdot   &\cdot  &1
\end{array}
\right),
\end{aligned}
$$
belongs to the face $\Phi[2,b,c]^\prime$ of $\ppt$,
where
$$
\la\mu=1,\qquad \lambda>0,\qquad \mu>0,\qquad \lambda\neq 1.
$$
This gives us a family of entangled states with positive partial
transposes. We show that these entangled states generate extreme
rays in the cone  $\ppt$ by showing that $A^{\prime\prime}$
consists of scalar multiples of $A$ with respect to the duality
between $\ppt$ and $\mathbb D$.

We note that
$$
A^\prime=\conv \{\phi_V+\phi^W: V,W\in D\}=\sigma(D,D)
$$
where $D$ denotes the orthogonal complement of $\spa \{x,y_1,y_2,y_3\}$, which is generated by
$\{V_1,V_2,V_3,W_1,W_2,W_3\}$.
If we write
$$
M=\left(\begin{matrix}1&\mu&-\la\\ -\la&1&\mu\\
\mu&-\la&1\end{matrix}\right)\qquad
L=\left(\begin{matrix}a_1&b_1&b_3\\ b_1&a_2&b_2\\
b_3&b_2&a_3\end{matrix}\right)
$$
then we see that an arbitrary element  $V$ of $D$ is of the form
$$
V=M\circ L
$$
with $a_1+a_2+a_3=0$, where $M\circ L$ denotes the Hadamard or
Schur product whose entries are obtained by the product of the
corresponding entries. From now on, the matrix representation
(\ref{choi-rep}) of a linear map $\phi$ will be denoted by the
same notation $\phi$. By a direct calculation, we have
$$
\phi^V=\phi^M\circ\phi^L
$$
with
$$
\phi^M=
\left(
\begin{array}{ccccccccccc}
1     &\mu   &-\la  &&-\la  &-1    &\la^2 &&\mu   &\mu^2 &-1    \\
\mu   &\mu^2 &-1    &&1     &\mu   &-\la  &&-\la  &-1    &\la^2 \\
-\la  &-1    &\la^2 &&\mu   &\mu^2 &-1    &&1     &\mu   &-\la  \\
\\
-\la   &1    &\mu   &&\la^2 &-\la   &-1    &&-1    &\mu   &\mu^2 \\
-1    &\mu   &\mu^2 &&-\la   &1    &\mu   &&\la^2 &-\la  &-1    \\
\la^2 &-\la  &-1    &&-1    &\mu   &\mu^2 &&-\la  &1     &\mu   \\
\\
\mu   &-\la  &1     &&-1    &\la^2 &-\la  &&\mu^2 &-1    &\mu   \\
\mu^2 &-1    &\mu   &&\mu   &-\la  &1     &&-1    &\la^2 &-\la  \\
-1    &\la^2 &-\la  &&\mu^2 &-1      &\mu   &&\mu   &-\la  &1
\end{array}
\right)
$$
and
$$
\phi^L= \left(
\begin{array}{ccccccccccc}
|a_1|^2     &\ol a_1b_1  &\ol a_1b_3  &&\ol b_1a_1  &|b_1|^2     &\ol b_1b_3  &&\ol b_3a_1  &\ol b_3b_1  &|b_3|^2     \\
\ol b_1a_1  &|b_1|^2     &\ol b_1b_3  &&\ol a_2a_1  &\ol a_2b_1  &\ol a_2b_3  &&\ol b_2a_1  &\ol b_2b_1  &\ol b_2b_3  \\
\ol b_3a_1  &\ol b_3b_1  &|b_3|^2     &&\ol b_2a_1  &\ol b_2b_1  &\ol b_2b_3  &&\ol a_3a_1  &\ol a_3b_1  &\ol a_3b_3  \\
\\
\ol a_1b_1  &\ol a_1a_2  &\ol a_1b_2  &&|b_1|^2     &\ol b_1a_2  &\ol b_1b_2  &&\ol b_3b_1  &\ol b_3a_2  &\ol b_3b_2  \\
|b_1|^2     &\ol b_1a_2  &\ol b_1b_2  &&\ol a_2b_1  &|a_2|^2     &\ol a_2b_2  &&\ol b_2b_1  &\ol b_2a_2  &|b_2|^2     \\
\ol b_3b_1  &\ol b_3a_2  &\ol b_3b_2  &&\ol b_2b_1  &\ol b_2a_2  &|b_2|^2     &&\ol a_3b_1  &\ol a_3a_2  &\ol a_3b_2  \\
\\
\ol a_1b_3  &\ol a_1b_2  &\ol a_1a_3  &&\ol b_1b_3  &\ol b_1b_2  &\ol b_1a_3  &&|b_3|^2     &\ol b_3b_2  &\ol b_3a_3  \\
\ol b_1b_3  &\ol b_1b_2  &\ol b_1a_3  &&\ol a_2b_3  &\ol a_2b_2  &\ol a_2a_3  &&\ol b_2b_3  &|b_2|^2     &\ol b_2a_3  \\
|b_3|^2     &\ol b_3b_2  &\ol b_3a_3  &&\ol b_2b_3  &|b_2|^2     &\ol b_2a_3  &&\ol a_3b_3  &\ol a_3b_2  &|a_3|^2
\end{array}
\right).
$$

Now, assume that a matrix $X=\sum_i\wt x_i\wt x_i^*\in\mathbb V_3\cap\mathbb V^3$ belongs to $A^{\prime\prime}$. Then
$X\in(\Phi_D)^\prime$ implies that $x_i\perp D$ for each $i$. Therefore, we see that $x_i$ is of the form
$$
x_i=\xi_i x+\alpha_i y_1+\beta_i y_2+\gamma_i y_3
=\rho\circ\sigma_i
$$
where
$$
\rho=
\left(\begin{matrix} 1&\la&\mu\\ \mu&1&\la\\ \la&\mu &1\end{matrix}\right)\quad
\sigma_i=
\left(\begin{matrix} \xi_i&\alpha_i&\gamma_i\\ \alpha_i&\xi_i&\beta_i\\ \gamma_i&\beta_i&\xi_i\end{matrix}\right).
$$
Therefore, it follows that
$$
X=\sum (\wt\rho \wt\rho^*)\circ(\wt\sigma_i \wt\sigma_i^*)
=(\wt\rho \wt\rho^*)\circ Y
$$
with
$$
\wt\rho\wt\rho^*=
\left(
\begin{array}{ccccccccccc}
1     &\la   &\mu  &&\mu  &1     &\la   &&\la   &\mu  &1     \\
\la   &\la^2 &1    &&1    &\la   &\la^2 &&\la^2 &1    &\la   \\
\mu  &1    &\mu^2 &&\mu^2 &\mu  &1    &&1    &\mu^2 &\mu  \\
\\
\mu  &1    &\mu^2 &&\mu^2 &\mu  &1    &&1    &\mu^2 &\mu  \\
1     &\la   &\mu  &&\mu  &1     &\la   &&\la   &\mu  &1     \\
\la   &\la^2 &1    &&1    &\la   &\la^2 &&\la^2 &1    &\la   \\
\\
\la   &\la^2 &1    &&1    &\la   &\la^2 &&\la^2 &1    &\la   \\
\mu  &1    &\mu^2 &&\mu^2 &\mu  &1    &&1    &\mu^2 &\mu  \\
1     &\la   &\mu  &&\mu  &1     &\la   &&\la   &\mu  &1
\end{array}
\right),
$$
and $Y= \sum \wt\sigma_i\wt\sigma_i^*$ is given by
$$
\left(
\begin{array}{ccccccccccc}
(\xi|\xi) &(\xi|\alpha)  &(\xi|\gamma) &&(\xi|\alpha)  &(\xi|\xi) &(\xi|\beta) &&(\xi|\gamma) &(\xi|\beta) &(\xi|\xi) \\
(\alpha|\xi)  &(\alpha|\alpha)   &(\alpha|\gamma)  &&(\alpha|\alpha)   &(\alpha|\xi)  &(\alpha|\beta)  &&(\alpha|\gamma)  &(\alpha|\beta)  &(\alpha|\xi)  \\
(\gamma|\xi) &(\gamma|\alpha)  &(\gamma|\gamma) &&(\gamma|\alpha)  &(\gamma|\xi) &(\gamma|\beta) &&(\gamma|\gamma) &(\gamma|\beta) &(\gamma|\xi) \\
\\
(\alpha|\xi)  &(\alpha|\alpha)   &(\alpha|\gamma)  &&(\alpha|\alpha)   &(\alpha|\xi)  &(\alpha|\beta)  &&(\alpha|\gamma)  &(\alpha|\beta)  &(\alpha|\xi)  \\
(\xi|\xi) &(\xi|\alpha)  &(\xi|\gamma) &&(\xi|\alpha)  &(\xi|\xi) &(\xi|\beta) &&(\xi|\gamma) &(\xi|\beta) &(\xi|\xi) \\
(\beta|\xi) &(\beta|\alpha)  &(\beta|\gamma) &&(\beta|\alpha)  &(\beta|\xi) &(\beta|\beta) &&(\beta|\gamma) &(\beta|\beta) &(\beta|\xi) \\
\\
(\gamma|\xi) &(\gamma|\alpha)  &(\gamma|\gamma) &&(\gamma|\alpha)  &(\gamma|\xi) &(\gamma|\beta) &&(\gamma|\gamma) &(\gamma|\beta) &(\gamma|\xi) \\
(\beta|\xi) &(\beta|\alpha)  &(\beta|\gamma) &&(\beta|\alpha)  &(\beta|\xi) &(\beta|\beta) &&(\beta|\gamma) &(\beta|\beta) &(\beta|\xi) \\
(\xi|\xi) &(\xi|\alpha)  &(\xi|\gamma) &&(\xi|\alpha)  &(\xi|\xi) &(\xi|\beta) &&(\xi|\gamma) &(\xi|\beta) &(\xi|\xi) \\
\end{array}
\right)
$$
if we denote by $\xi,\alpha,\beta$ and $\gamma$ the vectors
whose entries are $\xi_i,\alpha_i,\beta_i$ and $\gamma_i$,
respectively.

Now, we use the condition $X\in(\Phi^D)^\prime$ to see that
$$
\lan Y,(\wt\rho\wt\rho^*\circ\phi^M)\circ\phi^L\ran
=\lan\wt\rho\wt\rho^*\circ Y,\phi^M\circ\phi^L\ran =\lan
X,\phi^V\ran=0
$$
for every $a_1,a_2,a_3,b_1,b_2,b_3$ with $a_1+a_2+a_3=0$.
Note that
$$
\wt\rho\wt\rho^*\circ\phi^M=
\left(
\begin{array}{ccccccccccc}
+    &+    &-       &&-    &-    &\la^3    &&+    &\mu^3    &-   \\
+    &+    &-       &&+    &+    &-\la^3    &&-\la^3&-       &\la^3\\
-    &-    &+       &&\mu^3&\mu^3&-        &&+    &\mu^3    &-    \\
\\
-    &+    &\mu^3   &&+    &-    &-        &&-    &\mu^3    &\mu^3\\
-    &+    &\mu^3   &&-    &+    &+        &&\la^3&-        &-    \\
\la^3&-\la^3&-      &&-    &+    &+        &&-\la^3&+       &+\\
\\
+    &-\la^3&+      &&-    &\la^3&-\la^3   &&+    &-       &+\\
\mu^3&-     &\mu^3  &&\mu^3&-    &+        &&-    &+       &-\\
-    &\la^3 &-      &&\mu^3&-    &+        &&+    &-       &+
\end{array}
\right)
$$
where $\pm$ means $\pm 1$, respectively.

If we put $(1,-1,0), (0,1,-1), (-1,0,1)$ in the place of $(a_1,a_2,a_3)$ and $b_i=0$, then we have
$$
(\xi|\xi)=(\alpha|\alpha)=(\beta|\beta)=(\gamma|\gamma).
$$
Next, we put
$$
(1,-1,0;\sqrt{-1},0,0),\qquad (0,1,-1;0,\sqrt{-1},0),\qquad (-1,0,1;0,0,\sqrt{-1})
$$
in the place of $(a_1,a_2,a_3;b_1,b_2,b_3)$ to get
$$
(\xi|\alpha)=(\alpha|\xi),\qquad (\xi|\beta)=(\beta|\xi),\qquad (\xi|\gamma)=(\gamma|\xi),
$$
respectively. Putting $(1,-1,0;0,0,1)$ and $(1,-1,0;0,0,\sqrt{-1})$, we get
$$
(\alpha|\beta)+(\beta|\alpha)=0,\qquad (\alpha|\beta)-(\beta|\alpha)=0
$$
using the condition $\lambda\neq\mu$, and so, we have $(\alpha|\beta)=0$.
We also have $(\beta|\gamma)=0$, if we put $(0,1,-1;1,0,0)$ and $(0,1,-1;\sqrt{-1},0,0)$. Similarly, we also
have $(\gamma|\alpha)=0$, with $(-1,0,1;0,1,0)$ and $(-1,0,1;0,\sqrt{-1},0)$ in the place of $(a_1,a_2,a_3; b_1,b_2, b_3)$

Now, we may assume that $(\xi|\xi)=1$, and so we see that the block transpose $X^\tau$ of the matrix $X$ is of the following form
$$
\left(
\begin{array}{ccccccccccc}
1 &\la(\xi|\alpha)  &\mu(\xi|\gamma) &&\mu(\alpha|\xi)  &1   &\cdot        &&\la(\gamma|\xi) &\cdot  &1           \\
\la(\alpha|\xi)  &\la^2   &\cdot     &&1 &\la(\xi|\alpha)  &\mu(\xi|\gamma)&&\mu(\beta|\xi) &\cdot  &\cdot         \\
\mu(\gamma|\xi) &\cdot  &\mu^2       &&\la(\beta|\xi) &\cdot  &\cdot       &&1 &\la(\xi|\alpha)  &\mu(\xi|\gamma)  \\
\\
\mu(\xi|\alpha)  &1 &\la(\xi|\beta) &&\mu^2&\mu(\alpha|\xi)  &\cdot        &&\cdot  &\la(\gamma|\xi) &\cdot   \\
1   &\la(\alpha|\xi)  &\cdot        &&\mu(\xi|\alpha)  &1 &\la(\xi|\beta) &&\cdot  &\mu(\beta|\xi) &1 \\
\cdot  &\mu(\gamma|\xi) &\cdot      &&\cdot  &\la(\beta|\xi) &\la^2       &&\mu(\xi|\alpha)  &1 &\la(\xi|\beta)         \\
\\
\la(\xi|\gamma) &\mu(\xi|\beta) &1  &&\cdot  &\cdot  &\mu(\alpha|\xi)     &&\la^2&\cdot &\la(\gamma|\xi)    \\
\cdot  &\cdot  &\la(\alpha|\xi)     &&\la(\xi|\gamma) &\mu(\xi|\beta) &1  &&\cdot &\mu^2 &\mu(\beta|\xi)     \\
1 &\cdot &\mu(\gamma|\xi)           &&\cdot &1 &\la(\beta|\xi)           &&\la(\xi|\gamma) &\mu(\xi|\beta) &1  \\
\end{array}
\right).
$$
Since $X\in\mathbb V^3$ or equivalently $X^\tau\in\mathbb V_3$, this matrix
must be positive semi-definite. The positive semi-definiteness of the $3\times 3$ diagonal submatrix
with $2,3,4$ rows and columns tells us that
$(\xi|\beta)=0$.
From the $6,7,8$ and $2,6,8$ submatrices, we also have $(\xi|\alpha)=0$ and $(\xi|\gamma)=0$.
This completes the proof that $A^{\prime\prime}$ consists of the scalar multiples of $A$, and so the entanglement $A$ generates an
extreme ray in the cone $\ppt$.

One of the useful methods to construct entangled states with
positive partial transposes is to use the notion of unextendible
product basis as was considered in \cite{upb-entan},
\cite{upb-entan-1}, \cite{upb-entan-2}. It is easy to see that the
four dimensional subspace $D^\perp$ of $M_3$ has no rank one
matrix. On the other hand, the five dimensional subspace $D$ has
only six following rank one matrices
$$
\begin{gathered}
\left(\begin{matrix} 1&\mu&0\\-\la&-1&0\\0&0&0\end{matrix}\right)\qquad
\left(\begin{matrix} 1&0&\mu\\0&0&0\\-\la&0&-1\end{matrix}\right)\qquad
\left(\begin{matrix} 0&0&0\\0&1&\mu\\0&-\la&-1\end{matrix}\right)\\
\left(\begin{matrix} -1&\mu&0\\-\la&1&0\\0&0&0\end{matrix}\right)\qquad\ \
\left(\begin{matrix} -1&0&\mu\\0&0&0\\-\la&0&1\end{matrix}\right)\qquad\ \
\left(\begin{matrix} 0&0&0\\0&-1&\mu\\0&-\la&1\end{matrix}\right)
\end{gathered}
$$
up to scalar multiplication, which span $D$. But, no five
of them are orthogonal, and so, we see that our example does
not come out from unextendible product basis.

Because $A$ is of rank four, it follows by \cite{sanpera} that $A$ has Schmidt rank two. In this final paragraph,
we construct explicitly $3\times 3$ matrices $z_i\in M_3$ such that
$$
A=\sum_i {\wt z_i}\,{\wt z_i}^*,\qquad {\text{\rm rank of}}\ z_i\le 2.
$$
We define vectors $x_i=(x_{i1},x_{i2},x_{i3},x_{i4}),
y_i=(y_{i1},y_{i2},y_{i3},y_{i4}) \in \mathbb C^4$ by
$$
\begin{array}{ll}
x_1 = \left(\dfrac 12, \dfrac 12, \dfrac 12, \dfrac 12\right),
&x_3 = \left(\dfrac 1{2\lambda},\dfrac 1{2\lambda},-\dfrac 1{2\lambda},-\dfrac 1{2\lambda}\right),\\
x_2 = (\lambda^3\zeta, \zeta,\lambda^3\zeta, \zeta),\phantom{\dfrac{\frac 11}1}
&x_4 = (\zeta,\lambda^3\zeta,-\zeta, -\lambda^3\zeta),\\
y_1 = x_1,\quad
y_2 = -x_2,
&y_3 = x_3,\quad
y_4 = -x_4,
\end{array}
$$
for $i=1,2,3,4$,
where $\zeta = \sqrt{\dfrac {\lambda^2}{2+2\lambda^6}}$.
We also define
$$
z_i=\begin{pmatrix}
x_{1i} & x_{2i} & x_{3i}\\ \dfrac {x_{2i}}{\lambda^2} & x_{1i} & x_{4i}\\
\lambda^2 x_{3i}& \dfrac{x_{4i}}{\lambda^2} & x_{1i}
\end{pmatrix},\qquad
z_{i+4}=\begin{pmatrix}
y_{1i} & y_{2i} & y_{3i}\\ \dfrac {y_{2i}}{\lambda^2} & y_{1i} & y_{4i}\\
\lambda^2 y_{3i}& \dfrac{y_{4i}}{\lambda^2} & y_{1i}
\end{pmatrix},
$$
for $i=1,2,3,4$. We see that the determinant of $z_i$ is zero for each $i=1,2,\dots,8$.
By a direct calculation, we also have
$$
A = \dfrac 12 \sum_{i=1}^8 z_iz_i^*.
$$
and so $A$ belongs to $\mathbb V_2$. Since $A\notin\mathbb V_1$, we conclude that
the entanglement $A$ has Schmidt number two.


\begin{thebibliography}{99}

\bibitem{byeon-kye} E.-S. Byeon and S.-H. Kye, \it Facial
structures for positive linear maps in the two dimensional matrix
algebra, \rm Positivity, \bf 6 \rm (2002), 369--380.

\bibitem{upb-entan}
C. H. Bennett, D. P. DiVincenzo, T. Mor, P. W. Shor, J. A. Smolin and B. M. Terhal,
\it Unextendible product bases and bound entanglement,
\rm Phys. Rev. Lett.
\bf 82
\rm (1999),
5385--5388.

\bibitem{cho-kye-lee} S.-J. Cho, S.-H. Kye and S. G. Lee, \it Generalized Choi maps in
3-dimensional matrix algebras, \rm Linear Alg. Appl. \bf171 \rm (1992), 213--224.

\bibitem{choi75-10}  M.-D. Choi, \it Completely positive linear
maps on complex matrices, \rm Linear Alg. Appl. \bf 10 \rm
(1975), 285--290.

\bibitem{choi75}  M.-D. Choi, \it Positive semidefinite
biquadratic forms, \rm Linear Alg. Appl. \bf 12 \rm (1975),  95--100.

\bibitem{choi-lam}  M.-D. Choi and T.-T. Lam, \it Extremal positive
semidefinite forms, \rm Math. Ann. \bf 231 \rm (1977),
1--18.

\bibitem{upb-entan-1}
D. P. DiVincenzo, T. Mor, P. W. Shor, J. A. Smolin and B. M. Terhal,
\it Unextendible product bases, uncompletable product bases and bound entanglement,
\rm Commun. Math. Phys., to appear. 

\bibitem{upb-entan-2}
D. P. DiVincenzo and B. M. Terhal,
\it Product bases in quantum information theory,
\rm XIIIth International Congress on Mathematical Physics (London, 2000),
399--407,
Int. Press, Boston, 2001.

\bibitem{eom-kye} M.-H. Eom and S.-H. Kye, \it Duality for positive linear maps in matrix
algebras, \rm Math. Scand. \bf 86 \rm (2000), 130--142.

\bibitem{ha} K.-C. Ha, \it Atomic positive linear maps in matrix algebras, \rm
Publ. RIMS, Kyoto Univ.,  \bf 34 \rm (1998), 591--599.

\bibitem{ha-1} K.-C. Ha,
\it Positive projections onto spin factors,
\rm Linear Algebra Appl.
\bf 348
\rm (2002),
105--113.

\bibitem{horo-1}
M. Horodecki, P. Horodecki and R. Horodecki,
\it Separability of mixed states: necessary and sufficient conditions,
\rm Phys. Lett. A
\bf 223
\rm (1996),
1--8.

\bibitem{horo} M. Horodecki, P. Horodecki and R. Horodecki, \it Mixed-state entanglement and
quantum communication, \rm Quantum Information: An Introduction to Basic Theoretical
Concepts and Experiments, edited by G. Alber, et al, Springer Tracts in Modern Physics,
Vol. 173, pp. 151--195, Springer-Verlag, 2001.

\bibitem{horo-2}
P. Horodecki,
\it Separability criterion and inseparable mixed states with positive partial transposition,
\rm Phys. Lett. A
\bf 232
\rm (1997),
333--339.

\bibitem{itoh}
T. Itoh, \it Positive maps and cones in $C^*$-algebras, \rm Math.
Japonica \bf 31 \rm (1986), 607--616.

\bibitem{kim-kye}  H.-J. Kim and S.-H. Kye, \it Indecomposable
extreme positive linear maps in matrix algebras, \rm Bull. London
Math. Soc. \bf 26 \rm (1994),  575--581.

\bibitem{kraus} K. Kraus, \it Operations and effects in the Hilbert space
formulation of quantum theory, \rm Foundations of quantum mechanics and ordered
linear spaces (Marburg, 1973), pp. 206--229. Lecture Notes
in Phys., Vol. 29,  Springer, 1974.

\bibitem{kye-22} S.-H. Kye, \it A class of atomic positive maps in
$3$-dimensional matrix algebras,
\rm Elementary operators and applications (Blaubeuren, 1991),
pp. 205--209, World-Scientific, 1992.

\bibitem{kye-canad}  S.-H. Kye, \it Facial structures for positive
linear maps between matrix algebras, \rm Canad. Math. Bull. \bf 39 \rm (1996), 74--82.

\bibitem{kye-korean} S.-H. Kye, \it Boundaries of the cone of positive
linear maps and its subcones in matrix algebras, \rm J. Korean
Math. Soc. \bf 33 \rm (1996), 669--677.

\bibitem{kye-cambridge} S.-H. Kye, \it On the convex
set of all completely positive linear maps in matrix algebras, \rm
Math. Proc. Cambridge Philos. Soc. \bf 122 \rm (1997), 45--54.

\bibitem{kye-2by2_II} S.-H. Kye,
\it Facial structures for unital positive Linear Maps in the two dimensional matrix algebra,
\rm Linear Alg, Appl., \bf 362 \rm (2003), 57-- 73.

\bibitem{kye-decom} S.-H. Kye, \it Facial structures for
decomposable positive linear maps in matrix algebras, \rm
Positivity, to appear. [avaible at {\tt http://www.math.snu.ac.kr/$\sim$kye/paper.html}]

\bibitem{osaka}  H. Osaka, \it A class of extremal positive maps in
$3\times 3$ matrix algebras, \rm Publ. RIMS, Kyoto Univ. \bf 28 \rm
(1992),  747--756.

\bibitem{robertson}   A. G. Robertson, \it Positive projections on
$C^*$-algebras and extremal positive maps, \rm J. London Math. Soc.
(2) \bf 32 \rm (1985),  133--140.

\bibitem{sanpera}
A. Sanpera, D. Bru\ss\ and M. Lewenstein,
\it Schmidt number witnesses and bound entanglement,
\rm Phys. Rev. A \bf 63 \rm (2001), 050301.


\bibitem{stormer}  E. St\o rmer, \it Positive linear maps of operator
algebras \rm Acta Math. \bf 110 \rm (1963),  233--278.


\bibitem{stormer82}  E. St\o rmer, \it Decomposable positive maps on
$C^*$-algebras, \rm Proc. Amer. Math. Soc. \bf 86 \rm (1982),
402--404.

\bibitem{tomiyama} K. Tanahashi and J. Tomiyama, \it Indecomposable
positive maps in matrix algebras, \rm Canad. Math. Bull. \bf 31 \rm
(1988), 308--317.

\bibitem{tang}
W.-S. Tang, \it On positive linear maps between matrix algebras,
\rm Linear Algebra Appl. \bf 79 \rm (1986), 33--44.

\bibitem{terhal}
B. M. Terhal,
\it A family of indecomposable positive linear maps based on
            entangled quantum states,
\rm Linear Algebra Appl.
\bf 323
\rm (2001),
61--73.

\bibitem{terhal_horo}
B. M. Terhal and P. Horodecki,
\it Schmidt number for density matrices,
\rm Phys. Rev. A (3)
\bf 61
\rm (2000),
040301.

\bibitem{woronowicz}  S. L. Woronowicz, \it Positive maps of low
dimensional matrix algebras, \rm Rep. Math. Phys. \bf 10 \rm (1976),
165--183.


\end{thebibliography}
\end{document}